\documentclass[10pt, conference, letterpaper]{IEEEtran}
\IEEEoverridecommandlockouts
\usepackage[utf8]{inputenc}
\usepackage{tabularx,tabulary, adjustbox}
\usepackage{xcolor,tcolorbox,colortbl}
\usepackage{soul}
\usepackage{amsmath}
\usepackage[misc,clock,geometry]{ifsym}
\usepackage{amsthm,amssymb,graphicx,multirow,color,amsfonts}
\usepackage{caption, subcaption, cite, epstopdf, comment, mathtools}
\usepackage{subcaption}
\usepackage{mathrsfs}
\usepackage{pifont, mdsymbol}
\usepackage{breakurl}
\usepackage{array}
\usepackage{booktabs}
\usepackage{multirow}

\usepackage{amssymb}
\usepackage{amsmath}
\usepackage{amsthm}
\usepackage{amsfonts}
\usepackage{float}
\usepackage{placeins}

\DeclareMathAlphabet{\mathcal}{OMS}{cmsy}{m}{n}

\newcolumntype{C}[1]{>{\centering\arraybackslash}m{#1}}
\newcolumntype{Z}{>{\raggedright\let\newline\\\arraybackslash\hspace{0pt}}X}

\newcommand{\cut}[1]{}

\newcommand{\subparagraph}{}
\usepackage{titlesec}

\long\def\/*#1*/{}

\usepackage{fancyhdr}
\fancypagestyle{firstpagestyle}{
\fancyhf{}
\lhead{\footnotesize This is the author's accepted manuscript of the paper published in the IEEE Conference on Network Function Virtualization and Software-Defined Networking (NFV-SDN), 2025. Please cite the published IEEE version when referencing this work.
https://doi.org/10.1109/NFV-SDN66355.2025.11349452
}

\lfoot{\footnotesize © 2025 IEEE. Personal use of this material is permitted. Permission from IEEE must be obtained for all other uses, in any current or future media, including reprinting/republishing this material for advertising or promotional purposes, creating new collective works, for resale or redistribution to servers or lists, or reuse of any copyrighted component of this work in other works.
}

}

\def\BibTeX{{\rm B\kern-.05em{\sc i\kern-.025em b}\kern-.08em
    T\kern-.1667em\lower.7ex\hbox{E}\kern-.125emX}}

\usepackage[acronym]{glossaries}
\newacronym{5g}{5G}{Fifth Generation}
\newacronym{ru}{RU}{Radio Unit}
\newacronym{cu}{CU}{Centralized Unit}
\newacronym{du}{DU}{Distributed Unit}
\newacronym{ric}{RIC}{RAN Intelligent Controller}
\newacronym{ran}{RAN}{Radio Access Network}
\newacronym{kpi}{KPI}{Key Performance Indicator}
\newacronym{e2sm}{E2SM}{E2 service model}
\newacronym{ue}{UE}{User Equipment}
\newacronym{cm}{CM}{conflict mitigation}
\newacronym{cmf}{CMF}{conflict mitigation framework}
\newacronym{cd}{CD}{conflict detection}
\newacronym{cr}{CR}{conflict resolution}
\newacronym{ca}{CA}{conflict avoidance}
\newacronym{rl}{RL}{reinforcement learning}
\newacronym{ml}{ML}{machine learning}
\newacronym{drl}{DRL}{Deep Reinforcement Learning}
\newacronym{rcp}{RCP}{RAN Control Parameter}
\newacronym{gnn}{GNN}{graph neural network}
\newacronym{gcn}{GCN}{graph convolutional network}
\newacronym{ecdf}{ECDF}{empirical cumulative distribution function}
\newacronym{bler}{BLER}{block error rate}
\newacronym{prb}{PRB}{physical resource block}
\newacronym{mno}{MNO}{Mobile Network Operator}
\newacronym{dag}{DAG}{directed acyclic graph}
\newacronym{ate}{ATE}{Average Treatment Effect}
\newacronym{cate}{CATE}{Conditional Average Treatment Effect}

\makeglossaries

\begin{document}
\setacronymstyle{long-short}


\title{Towards xApp Conflict Evaluation with Explainable Machine Learning and Causal Inference in O-RAN}

\author{
\IEEEauthorblockN{Pragya Sharma\IEEEauthorrefmark{1},
Shihua Sun\IEEEauthorrefmark{1},
Shachi Deshpande\IEEEauthorrefmark{2},
Angelos Stavrou\IEEEauthorrefmark{1},
Haining Wang\IEEEauthorrefmark{1}}
\IEEEauthorblockA{\IEEEauthorrefmark{1}Department of Electrical and Computer Engineering, Virginia Tech, USA\\
\IEEEauthorrefmark{2}Department of Computer Science, Cornell University, USA\\
Email: \IEEEauthorrefmark{1}\{pragyasharma, shihuas, angelos, hnw\}@vt.edu, \IEEEauthorrefmark{2}shachi@cs.cornell.edu
}
}

\maketitle
\thispagestyle{firstpagestyle}

\begin{abstract}

The Open Radio Access Network (O-RAN) architecture enables a flexible, vendor-neutral deployment of 5G networks by disaggregating base station components and supporting third-party xApps for near real-time RAN control. However, the concurrent operation of multiple xApps can lead to conflicting control actions, which may cause network performance degradation. In this work, we propose a framework for xApp conflict management that combines explainable machine learning and causal inference to evaluate the causal relationships between RAN Control Parameters (RCPs) and Key Performance Indicators (KPIs). We use model explainability tools such as SHAP to identify RCPs that jointly affect the same KPI, signaling potential conflicts, and represent these interactions as a causal Directed Acyclic Graph (DAG). We then estimate the causal impact of each of these RCPs on their associated KPIs using metrics such as Average Treatment Effect (ATE) and Conditional Average Treatment Effect (CATE). This approach offers network operators guided insights into identifying conflicts and quantifying their impacts, enabling more informed and effective conflict resolution strategies across diverse xApp deployments.

\end{abstract}


\IEEEpubidadjcol

\section{Introduction}
\glsunset{ran}

\begin{figure}[t!]
    \centering
    \includegraphics[width=0.82\columnwidth, trim={2.1cm 2.8cm 2.1cm 2cm},clip]{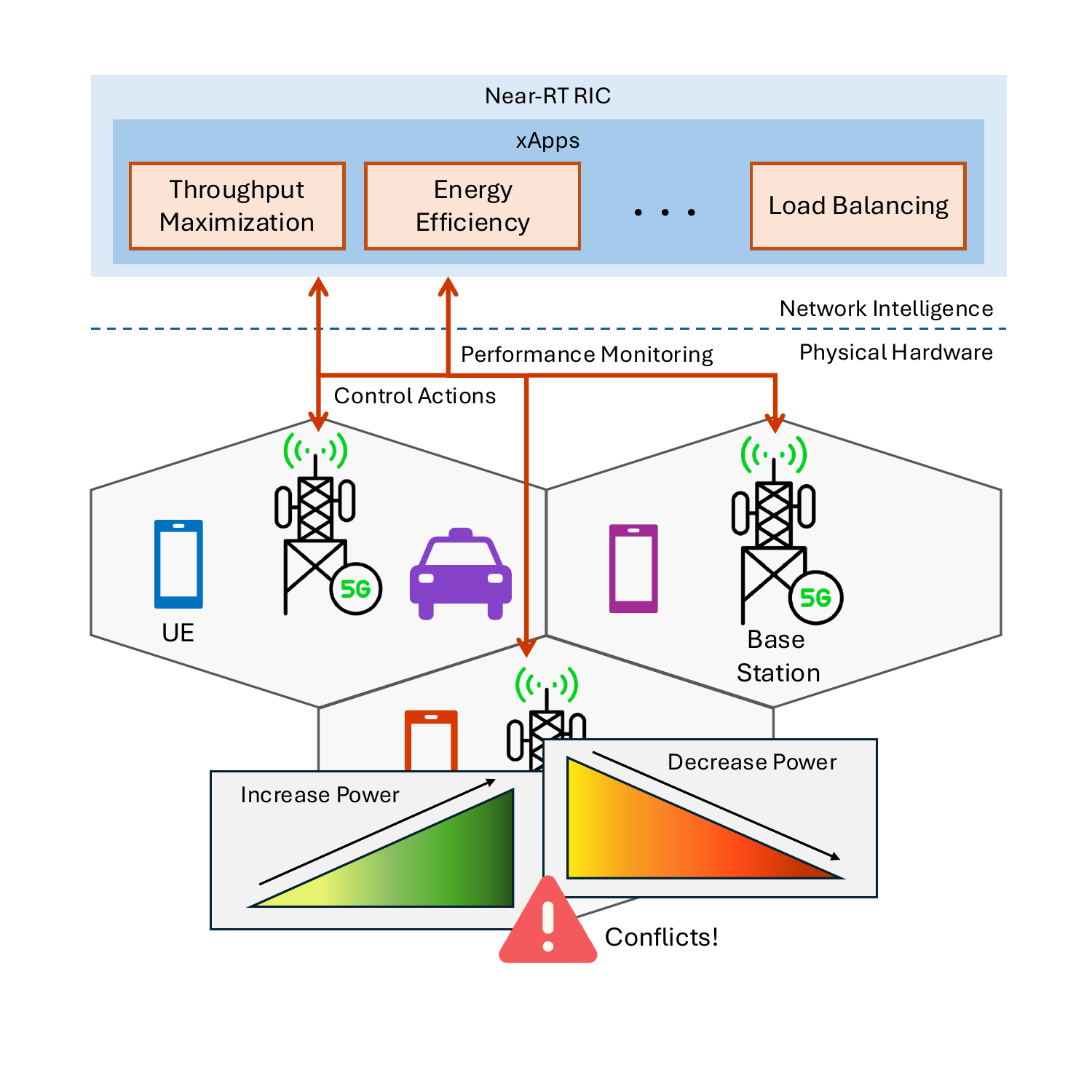}
     \caption{Example of potential conflicting interactions between xApps with different objectives.}
     \vspace{-2mm}
     \label{fig:intro}
\end{figure}

The Open Radio Access Network (O-RAN) Alliance is reshaping mobile network deployments by fostering a multi-vendor ecosystem with open interfaces to ensure interoperability among RAN components. 
The O-RAN architecture~\cite{oran-wg1-arch} disaggregates 5G base stations (gNBs) into \glspl{cu}, \glspl{du}, and \glspl{ru}, each implementing different layers of the RAN protocol stack. This modular design enables \glspl{mno} to flexibly deploy and scale components based on demand. These components are managed by two software-defined controllers known as the \glspl{ric}, both hosting third-party applications. The near Real-Time RIC (near-RT RIC) hosts xApps for time-sensitive tasks (10-1000 ms), while the non Real-Time RIC (non-RT RIC) hosts rApps for longer timescale operations ($>$1000 ms), such as analytics and AI/ML model training and inference.

xApps function as independent agents that manage RAN control operations such as network slicing, traffic steering, and energy efficiency. In deployments with multiple xApps, the actions of one xApp may unintentionally interfere with others, resulting in conflicts as depicted in Figure~\ref{fig:intro}. These conflicts can degrade performance and disrupt network operations. Although the O-RAN Alliance acknowledges the risks associated with such interactions~\cite{oran-wg3-confmit}, no standardized mechanisms currently exist to detect or mitigate xApp conflicts.

Research on xApp conflict management is still in the nascent stage. Existing efforts span from game-theoretic methods~\cite{wadud2025,WadudQACM} to \gls{ml}-based techniques, including \gls{rl}~\cite{ZhangICC2022,ZhangSensors,xappDistillation} and \glspl{gnn}~\cite{santos2025,shami2025}. While these works provide valuable insights for specific use cases, no existing work provides a comprehensive framework to identify and assess the impact of conflicting \glspl{rcp} on network \glspl{kpi} applicable across diverse xApp deployments.

In this study, we aim to address these critical gaps by answering two primary questions: 1) How can conflicting RCPs be systematically identified, particularly when they jointly influence the same KPI? and 2) How can we quantify the relative impact of each conflicting RCP on the affected KPI, especially considering variations in network states and operational conditions? To tackle these research questions, we introduce a novel conflict evaluation framework that integrates explainable \gls{ml} and causal inference techniques within the O-RAN ecosystem for robust xApp conflict management. We begin by applying SHAP~\cite{shap}, an \gls{ml} explainability method, to a regression model trained on network observational data to identify RCPs that may cause potential conflicts. Subsequently, we adopt a rigorous causal effect estimation approach~\cite{Pearl_2009,causalSurvey} to quantify the impact of individual RCPs on corresponding KPIs, accounting for dynamic changes in network states. 

Our proposed approach equips network operators with a robust tool for understanding and managing trade-offs between conflicting xApps, facilitating the development of more informed and effective conflict resolution policies.  Furthermore, with its foundation in causal analysis, the solution is agnostic to specific xApp functionalities, as it focuses on the underlying cause-and-effect relationships between \glspl{rcp} and \glspl{kpi}.
To the best of our knowledge, this is the first work to comprehensively integrate \gls{ml} model explainability and causal inference into the xApp conflict management pipeline within O-RAN. 

The remainder of this paper is organized as follows. Section~\ref{sec:relatedWork} presents a comprehensive review of existing literature on xApp conflict management. Section~\ref{sec:conflictModeling} introduces the modeling framework for characterizing xApp conflicts. In Section~\ref{sec:methodology}, we describe our proposed methodology for conflict identification and impact estimation. Section~\ref{sec:perfEval} details the evaluation of our approach using simulated network data. Finally, we discuss the limitations of our current evaluation and outline directions for future work in Section~\ref{sec:discussion}, followed by a summary of the paper in Section~\ref{sec:conclusion}.

\section{Related Work} \label{sec:relatedWork}

The O-RAN Alliance recently released a technical report on conflict mitigation among xApps in the near-RT RIC~\cite{oran-wg3-confmit}, categorizing conflicts into three types: direct, indirect, and implicit (see Section~\ref{sec:conflictModeling} for formal definitions and examples). While the report is a valuable foundation, no standardized methods currently exist for managing such conflicts, thus prompting active research in this area.

Some existing works propose enabling cooperation among xApps to prevent conflicts from occurring. In~\cite{ZhangICC2022, ZhangSensors}, a team learning-based framework is introduced where xApps act as cooperative \gls{drl} agents, sharing their intended actions with each other during training. However, this approach assumes collaboration among third-party vendors during xApp development, which may not be realistic in commercial O-RAN deployments. 
\cite{xappDistillation} introduces a knowledge distillation technique that merges multiple \gls{drl}-trained xApps into a unified xApp, which learns to select the most effective action based on combined policies. It remains unclear how this approach can be extended to xApps operating with different state and action spaces. 

Policy-driven conflict resolution is also discussed in some works. COMIX~\cite{comix} presents a conflict mitigation framework for resolving direct conflicts between \gls{drl}-based power control xApps by evaluating various conflict resolution policies and their trade-offs between power consumption and throughput. However, this method is limited to direct conflicts within a specific class of xApps. In~\cite{xappConflictScheduler}, a scheduler-based conflict mitigation strategy based on the Advantage Actor-Critic (A2C) method is proposed to determine which xApps are allowed to operate. The authors in~\cite{AdamczykInfocomWkshp, AdamczykIEEEComm2023} present a priority-based conflict resolution framework that resolves conflicts by enforcing a static priority order among xApps. This approach, however, lacks guidance on how \glspl{mno} should assign optimal priorities, especially in dynamic environments where xApp goals may change over time.

The authors in~\cite{WadudQACM,wadud2025} propose a game-theoretic approach to conflict mitigation, where xApps reach consensus on RCP values to minimize KPI degradation due to conflicts. However, their methodology presumes that the relationships among xApps, RCPs, and KPIs are known a-priori and adhere to a Gaussian distribution. In practical settings, these relationships are often difficult to discern and may lack consistent patterns due to the dynamic nature of network conditions, rendering such assumptions impractical.

Graph-based learning has also found traction in some recent studies. In~\cite{santos2025}, an unsupervised \gls{gnn}-based approach is presented to learn the hidden relationships among xApps, \glspl{rcp}, and \glspl{kpi}, enabling the reconstruction of conflict graphs. Like~\cite{WadudQACM}, their evaluation assumes Gaussian relationships between \glspl{rcp} and \glspl{kpi}, which may not reflect real-world complexity.\cite{shami2025} presents a \gls{gcn}-based framework for conflict classification using a manually labeled synthetic dataset with binary inputs for system states and integer outputs for conflict categories. This evaluation on synthetic, discretized data raises concerns regarding its applicability to real-world network scenarios.

The work most closely related to ours is PACIFISTA~\cite{pacifista}, which models the relationships between xApps, \glspl{rcp}, and \glspl{kpi} as hierarchical conflict graphs to identify and evaluate conflicts in O-RAN. It leverages a profiling pipeline for generation of statistical profiles of xApps using a network digital twin. However, the statistical profiles only include the cumulative distribution functions of different KPIs, which do not reveal more granular relationships between RCPs and KPIs across dynamic network states. In contrast, we utilize explainable \gls{ml} and causal inference techniques on raw network data to identify conflicting RCPs and measure their impact on KPIs.

\section{xApp Conflicts Definition and Modeling} \label{sec:conflictModeling}

A key aspect towards the overall conflict management of xApps, which includes conflict detection and resolution, is understanding the interdependent relationships between xApps, their \glspl{rcp}, and the \glspl{kpi}. Our proposed approach in this work is to analyze those relationships through a cause-and-effect lens, i.e., which xApps and by extension the \glspl{rcp} controlled by those xApps, are the causes of potential conflicts and how they impact network \glspl{kpi}.

To capture the interdependencies between xApps, RCPs, and network KPIs, we incorporate a graph-based representation that models these relationships and highlights potential conflicts. Let $\mathcal{A}$ be the set of xApps deployed in the network controlling the set $\mathcal{P}$ of \glspl{rcp} and observing the set $\mathcal{K}$ of \glspl{kpi}. As shown in Figure~\ref{fig:conflictGraphs}, the network can be represented as a directed graph $\mathcal{G} = (\mathcal{V},\mathcal{E})$ with vertices $\mathcal{V}= \mathcal{A} \cup \mathcal{P} \cup \mathcal{K}$ and set of edges $\mathcal{E}$. According to the O-RAN Alliance~\cite{oran-wg3-confmit}, xApp conflicts can be categorized into three types: 

\begin{figure}[t!]
    \centering
    \includegraphics[width=0.75\columnwidth,trim={2cm 4.5cm 9cm, 0.8cm},clip]{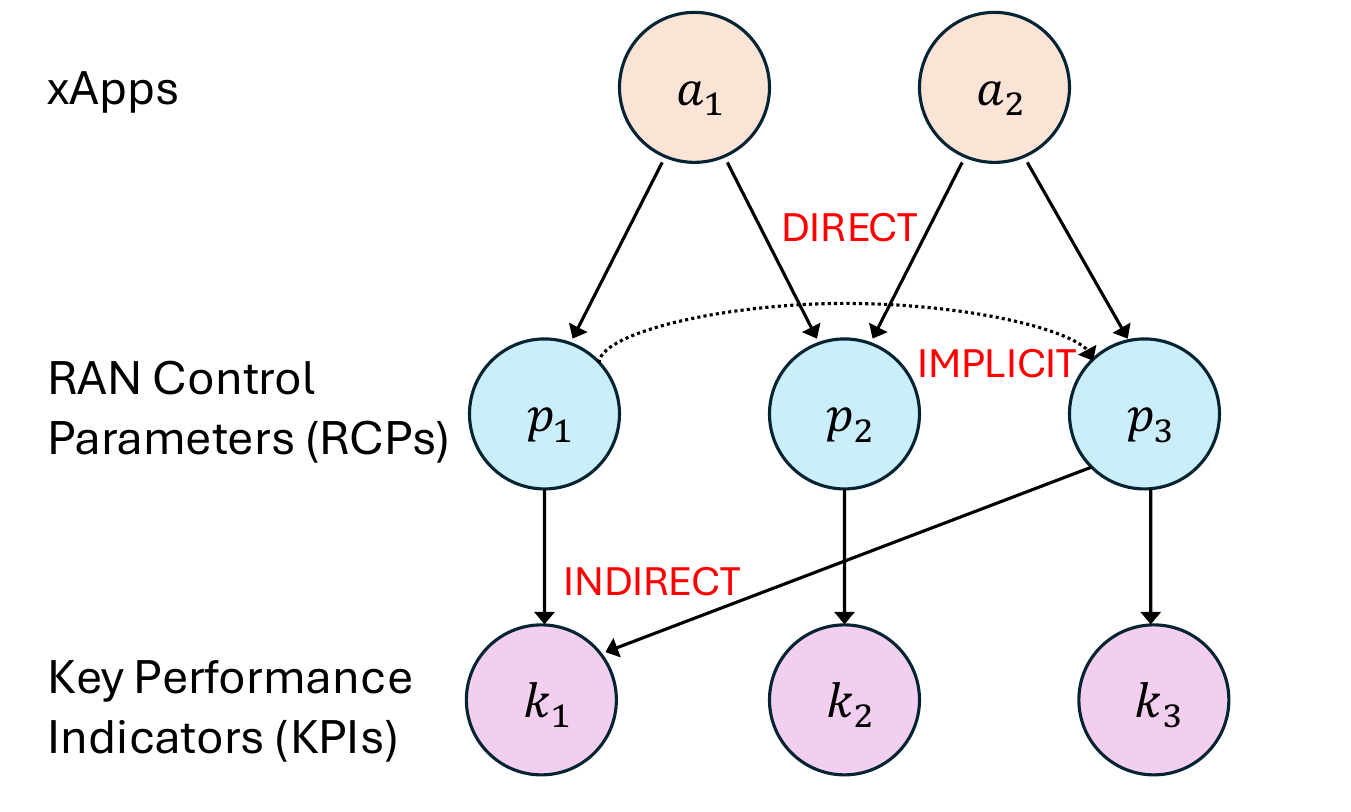}
     \caption{Directed graph with edges representing direct, indirect and implicit conflicts among xApps.}
     \label{fig:conflictGraphs}
\end{figure}

\begin{itemize}
    \item \textbf{Direct Conflicts} - Occur when two xApps $a_i, a_j \in \mathcal{A}$ control the same \gls{rcp} $p \in \mathcal{P}$ i.e $ a_i \rightarrow p, a_j \rightarrow p \in \mathcal{E} $. For instance, a RAN slice management xApp and a load balancing xApp may both modify the number of \glspl{prb} assigned to different slices. In Figure~\ref{fig:conflictGraphs}, xApps $a_1$ and $a_2$ conflict directly over \gls{rcp} $p_2$.

    \item \textbf{Indirect Conflicts} - Occur when two xApps $a_i, a_j \in \mathcal{A}$ control different \glspl{rcp} $p_m, p_n \in \mathcal{P}$ that influence the same \gls{kpi} $k \in \mathcal{K}$ i.e. $a_i \rightarrow p_m, a_j \rightarrow p_n,  p_m \rightarrow k, p_n \rightarrow k \in \mathcal{E} $. For instance, a resource allocation xApp adjusting PRBs and a power allocation xApp modifying transmit power can both impact downlink throughput. In Figure~\ref{fig:conflictGraphs}, the xApps $a_1$ and $a_2$ indirectly conflict through \glspl{rcp} $p_1$ and $p_3$ in affecting \gls{kpi} $k_1$. 
    
    \item \textbf{Implicit conflicts} - Occur when two xApps $a_i, a_j \in \mathcal{A}$ aim to optimize different \glspl{kpi} $k_y, k_z \in \mathcal{K}$ via different \glspl{rcp} $p_m, p_n \in \mathcal{P}$, but one \gls{rcp} implicitly influences the other, affecting its associated \gls{kpi}. Formally, if $a_i \rightarrow p_m, a_j \rightarrow p_n, p_m \rightarrow k_y, p_n \rightarrow k_z $ and $p_m \rightarrow p_n \in \mathcal{E}$, then there exists an implicit path $ p_m \rightarrow p_n \rightarrow k_z $. For example, a spectral efficiency xApp that adjusts cell bandwidth might interfere with the performance of a RAN slicing xApp as bandwidth influences \gls{prb} calculation. In Figure~\ref{fig:conflictGraphs}, the xApps $a_1$ and $a_2$ are in implicit conflict as \gls{rcp} $p_1$ modifies \gls{rcp} $p_3$, which in turn affects \gls{kpi} $k_3$.

\end{itemize}

\section{Conflict Detection and Impact Estimation} \label{sec:methodology}

In this section, we first provide the overview of our proposed approach for xApp conflict detection and evaluation, followed by detailed methodology describing the identification of conflicting RCPs through explainable \gls{ml} techniques. Finally, we discuss the estimation of impact of the identified conflicts on network performance using formal causal inference methods. 


\subsection{Motivation and Overview of Proposed Approach} \label{sec:overview}

As mentioned in section~\ref{sec:conflictModeling}, direct conflicts arise when multiple xApps control the same \gls{rcp}. Their detection is relatively straightforward by identifying common RCPs through xApp service model subscriptions and configuration data. In contrast, detecting and resolving indirect and implicit conflicts is significantly more complex, as it involves understanding how different xApps, controlling distinct \glspl{rcp}, influence shared \glspl{kpi}. 
For example, consider an xApp that adjusts cell bandwidth to improve spectral efficiency and another that modifies transmit power to enhance energy efficiency. While they operate on different \glspl{rcp}, their combined influence may degrade a shared \gls{kpi} such as average throughput. In such cases, it is critical to assess both the qualitative impact (e.g., identifying throughput degradation) and the quantitative impact (e.g., measuring the extent of degradation caused by a specific change in bandwidth or power). This dual understanding enables operators to define acceptable conflict tolerance thresholds in scenarios where preventing such conflicts entirely is not feasible.

\begin{figure}[t!]
    \centering
    \includegraphics[width=0.88\columnwidth,trim={2cm 1.7cm 6cm, 4.5cm},clip]{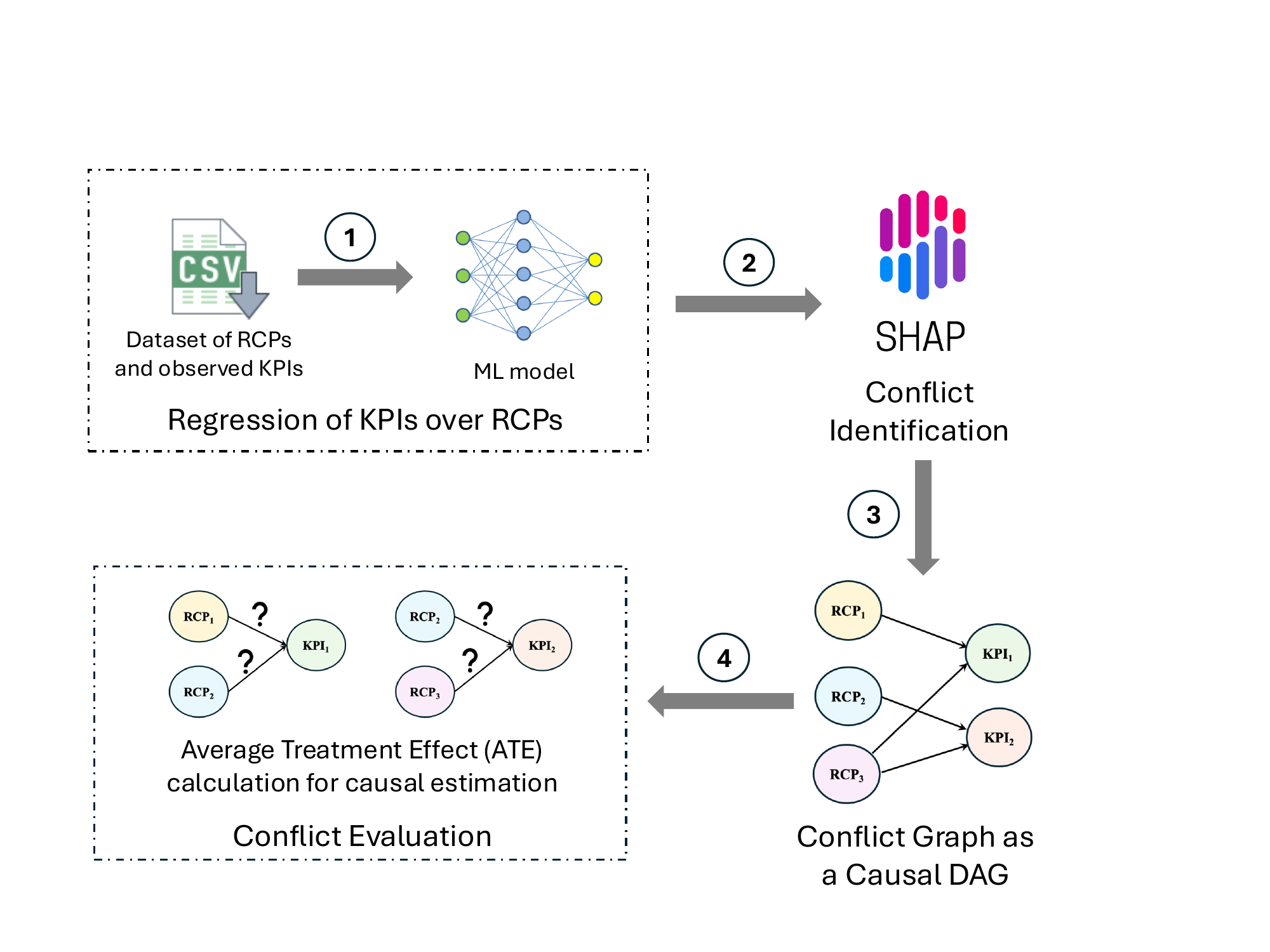}
     \caption{Proposed approach towards conflict evaluation}
     \vspace{-3mm}
     \label{fig:proposedMethod}
\end{figure}

To this end, in this work we focus on two aspects of evaluating xApp conflicts \textbf{(1) identifying RCPs that interfere with each other in influencing the same network \glspl{kpi}}, and \textbf{(2) quantifying the impact of these conflicting RCPs on the affected \glspl{kpi}.} 
As illustrated in Figure~\ref{fig:proposedMethod}, to address the first objective, \ding{192} we employ \acrfull{ml}-based regression to analyze the joint influence of multiple RCPs on KPIs, and \ding{193} utilize explainable ML techniques such as SHAP~\cite{shap} to determine the relative importance of each RCP in affecting the KPIs.
To achieve the second goal, \ding{194} we construct a causal \gls{dag} informed by the insights derived from the SHAP analysis, which allows us to \ding{195} estimate the numerical causal effect of each conflicting RCP on the KPIs. The subsections that follow provide a detailed explanation of each of these steps.

Finally, our proposed approach can be applied to xApps with different use cases, as it relies solely on the underlying causal relationships between RCPs and KPIs. This analytical framework is well-aligned with the considerations outlined in Clause 5.1.3 \textit{Key Issue \#3: Detection of E2-related Indirect and/or Implicit Conflicts} in the O-RAN Alliance's Conflict Mitigation Report~\cite{oran-wg3-confmit}. 

\subsection{Conflict Detection}\label{sec:confDet}

\subsubsection{Data Requirements}
The first step towards identifying conflicts among xApps in O-RAN is collecting a rich observational data that reflects network states across different operational scenarios. Such data can be obtained by experiments on a RAN simulator (e.g.,~\cite{nsORAN}) or a network digital twin (e.g.,~\cite{colosseum}). For the purpose of xApp conflict evaluations, we are interested specifically in a dataset that contains the values of \glspl{rcp} controlled by various xApps at the beginning of a time interval and the network \glspl{kpi} observed at the end of that time interval, across multiple periods of time. Such a dataset represents temporal snapshots of the network states, with RCPs serving as input features and KPIs as corresponding output targets. By systematically analyzing these RCP-KPI relationships, we aim to identify conflicting effects of RCPs on KPIs, effectively identifying conflicting actions across different xApps.

\vspace{1.5mm}
\subsubsection{ML Regression and Model Explainability}
While ML regression models are typically optimized to achieve accurate predictive performance, a robustly trained model can also serve as an analytical tool, providing meaningful insights into the relative importance of input features (RCPs) on outputs (KPIs). In our case, training an ML regression model on observational data like the one described above helps us understand the impacts of RCPs on KPIs and provides valuable insights into identifying which RCPs predominantly drive changes in network KPIs, thus helping network operators pinpoint critical areas of intervention.

Once a reliable machine learning model has been trained to regress KPIs on RCPs, we apply model explainability techniques, specifically SHAP (SHapley Additive exPlanations)~\cite{shap}, to identify which RCPs influence the same KPI, thereby revealing potential conflicts among xApps. SHAP uses cooperative game theory to assign each input feature (RCP) a numerical score that weighs its contribution to the model’s prediction of the KPI. Higher SHAP values indicate a stronger influence of a given RCP on the KPI. When multiple RCPs exhibit SHAP values of similar magnitude for a specific KPI, it suggests that they have comparable levels of influence on that KPI. Identifying these shared influences allows us to construct conflict graphs, such as the one shown in Figure~\ref{fig:conflictGraphs}.

While explainable ML techniques are effective in uncovering strong correlations between inputs and outputs, they do not inherently capture causal relationships. The objective of our work is to go beyond correlation and reveal the underlying cause-and-effect dynamics between xApp control actions (RCPs) and network performance metrics (KPIs). Although feature importance scores derived from explainability methods can guide the identification of potential conflicts, confirming whether the implicated RCPs are true causes of KPI degradation requires a formal causal inference framework, as discussed in the following section.

\subsection{Conflict Impact Estimation}\label{sec:confImpactEst}

Before we dive deeper into the application of causal inference methods for xApp conflict management, we first define the terminology used in causal analysis and its mapping to our system variables, namely RCPs and KPIs. In a causal inference framework~\cite{Pearl_2009, causalSurvey}, the variables whose effects are to be evaluated are referred to as \textit{treatments}, while the variables whose responses to those treatments are observed are termed \textit{outcomes}. Variables that influence both the treatment and the outcome are known as \textit{confounders}. Causal inference on observational data aims to estimate the effect of a treatment on an outcome while adjusting for all known confounders. In our context, RCPs are treated as the treatments and KPIs as the outcomes. In the presence of implicit conflicts, certain RCPs may also serve as confounders. By employing causal effect estimation techniques, we can quantify the impact of each RCP on the corresponding KPIs.

\vspace{1.5mm}
\subsubsection{Causal Directed Acyclic Graph (DAG) Construction.} 
To transform observational data into suitable information for a causal analysis model, the first step is to construct a causal \gls{dag}. The causal \gls{dag} represents assumptions of existing causal links between the treatments, confounders, and outcomes. Following the structure of Figure~\ref{fig:conflictGraphs}, the potential causal DAG can be seen as a subgraph $\mathcal{G}^{\prime} = (\mathcal{V}^{\prime}, \mathcal{E}^{\prime})$ consisting of nodes $ \mathcal{V}^\prime = \mathcal{P} \cup \mathcal{K}$. The set of edges $\mathcal{E}^\prime$ is the combined set of edges representing indirect and implicit conflicts and represents causal links in the graph $\mathcal{G}^{\prime}$. Note that the causal \gls{dag} would not include direct conflicts because those are represented by edges from xApp nodes $\mathcal{A}$ to \gls{rcp} nodes $\mathcal{P}$ and do not form part of subgraph $\mathcal{G}^{\prime}$. We construct this \gls{dag} by leveraging the SHAP analysis done in the previous subsection and creating causal edges from the RCPs which are the most influential towards their associated KPIs.


\vspace{1.5mm}
\subsubsection{Treatment Effect Estimation} 
To quantify the causal edges on the \gls{dag}, we compute two key metrics. First is the \textit{\gls{ate}}~\cite{Pearl_2009,causalSurvey}, which provides an estimated value representing the average impact of changing an \gls{rcp} on a \gls{kpi} across all conditions. Unlike a simple correlation, the ATE estimates the outcome of a hypothetical intervention, conceptually answering: `If we increase this \gls{rcp} by one unit, by how much can we expect the network \gls{kpi} to change on average?' 

According to conflict graph modeling in Section~\ref{sec:conflictModeling}, ATE estimation enables a straightforward evaluation of indirect conflicts where an RCP $p$ influences a KPI $k$ (i.e. $p\rightarrow k$). However, implicit conflicts such as $p_m \rightarrow p_n \rightarrow k_z$ manifest as `backdoor paths' in the causal \gls{dag} from the treatment RCP ($p_n$) to the outcome KPI ($k_z$). These paths arise due to confounding RCPs ($p_m$) and can lead to spurious correlations that bias the ATE estimates. To obtain an unbiased estimate of the true causal effect between a treatment-outcome pair, the estimation procedure must adjust for all identified confounders by conditioning on them, in accordance with Pearl’s backdoor criterion~\cite{Pearl_2009}.

To gain deeper insight beyond the ATE, we compute the \textit{\gls{cate}}~\cite{Pearl_2009,causalSurvey}, which measures the expected impact of a treatment under specific network conditions. Unlike a global average, the CATE answers the practical question: `Given the current network state, how much would the KPI change if we adjusted this RCP?' This captures heterogeneity in treatment effects across varying states, shifting the analysis from a single estimate to a distribution of conditional impacts. For MNOs, this granularity is especially valuable, as it reveals when specific parameter adjustments are most effective. Such targeted insights enable scenario-specific policy decisions to mitigate xApp conflicts, particularly in frequently occurring or critical network states, ultimately enhancing network performance.

\section{Evaluations} \label{sec:perfEval}
In this section, we describe the simulation setup and systematically evaluate our approach to answer the following research questions: 
\textbf{(RQ1)} - How to identify RCPs that conflict with each other in influencing the same KPI?
\textbf{(RQ2)} - How to quantify the impact of each conflicting RCP on the affected KPI? Furthermore, how does this impact vary across different observed network states?

\subsection{Network Simulation}

\begin{table}[b]
    \centering
    \caption{Simulation Parameters}
    \label{tab:simParam}        
    \begin{tabular}{|l|p{3cm}|}
        \toprule
        \textbf{Parameter} & \textbf{Value} \\
        \midrule 
        Carrier Frequency & 2.6 GHz \\
        Subcarrier Spacing & 15 kHz \\
        Duplex Mode & FDD \\
        Channel Model & Urban Micro \\
        Scheduler & Proportional Fair \\
        Transmit Power & [1,...., 40] dBm \\
        Bandwidth & [5, 10, 15, 20, 25, 30, 35, 40, 45, 50] MHz \\
        Number of PRBs & [4,...., PRB\textsubscript{max}] \\
        Number of Tx Antennas & [1,2,4,8,16] \\ 
        \bottomrule
    \end{tabular}
\end{table}

We design a system-level RAN simulation in MATLAB utilizing the 5G Toolbox and Communications Toolbox Wireless Network Simulation Library. The simulation scenario consists of a single gNB and one UE, both configured with a video conferencing traffic profile. Table~\ref{tab:simParam} outlines the simulation parameters. 
In this setup, the deployed xApps are assumed to control four \glspl{rcp}: bandwidth, number of PRBs, transmit (Tx) power, and number of Tx antennas. The bandwidth and number of Tx antennas are randomly sampled from discrete values provided in Table~\ref{tab:simParam}, while integer values for Tx power and number of PRBs are uniformly sampled from their respective ranges. The maximum number of PRBs (PRB\textsubscript{max}) in the sampling range is adjusted to comply with the standardized values in Table 5.3.2-1 in 3GPP TS38.104~\cite{3gppTS38104} according to the sampled bandwidth. The simulation is run for 1000 episodes, with each episode spanning 100 frames. The \glspl{kpi} measured are throughput (in Mbps), spectral efficiency, and \gls{bler}.

\subsection{Conflict Identification via Explainable ML Regression}

\begin{table*}[htb!]
\vspace{0.1in}
\centering
\caption{Performance metrics for regression of KPIs over RCPs across different models}
\begin{tabular}{|l|ccc|ccc|ccc|}
\toprule
\multirow{2}{*}{\textbf{Model}} & \multicolumn{3}{c|}{\textbf{Throughput}} & \multicolumn{3}{c|}{\textbf{Spectral Efficiency}} & \multicolumn{3}{c|}{\textbf{BLER}} \\ 
\cmidrule(lr){2-4} \cmidrule(lr){5-7} \cmidrule(lr){8-10}
 & R\textsuperscript{2} & MSE & MAE & R\textsuperscript{2} & MSE & MAE & R\textsuperscript{2} & MSE & MAE\\
\midrule

DecisionTree  &0.9221    &0.0381    &0.1522        &0.9567  &0.0005  &0.0133        &0.9596  &0.0042  &0.0412\\
RandomForest  &0.9329    &0.0328    &0.1365        &0.9633  &0.0004  &0.0109        &0.9756  &0.0025  &0.0334 \\
XGBoost       &\textbf{0.9364} &\textbf{0.0311}  &\textbf{0.1323}  &\textbf{0.9703}  &\textbf{0.0003}  &\textbf{0.0103} &\textbf{0.9866}  &\textbf{0.0014}  &\textbf{0.0268}\\
MLP           &0.9162    &0.0410    &0.1529        &0.9411  &0.0007  &0.0170        &0.9787  &0.0022  &0.0346\\
SVR           &0.8774    &0.0600    &0.1983        &0.9626  &0.0004  &0.0119        &0.9714  &0.0030  &0.0407\\
\bottomrule
\end{tabular}%

\label{tab:regression_metrics}
\end{table*}

\begin{table}[tb!]
    \centering
    \caption{Feature importance of various RCPs on KPIs: Throughput, Spectral Efficiency and BLER via (a) Permutation Importance (b) SHAP, for XGBoost regression model}

    \begin{subtable}{\columnwidth}
        \caption{Permutation Importance Scores}
        \vspace{-1mm}
        \resizebox{\columnwidth}{!}{%
            \begin{tabular}{|lccc|}
                \toprule
                \textbf{RCP} & \textbf{Throughput} & \textbf{Spectral Efficiency} & \textbf{BLER}\\
                \midrule
                Tx Power            &1.8445     &0.4134     &1.7116         \\
                Bandwidth           &0.0038     &1.4512     &0.0022         \\
                PRBs                &0.0175     &0.0095     &0.4918         \\
                Num of Tx Antenna   &0.0826     &0.0067     &0.0497         \\
                \bottomrule
            \end{tabular}
        }%
        \label{tab:permImpScores}
    \end{subtable}
    
    \vspace{2mm}
    \begin{subtable}{\columnwidth}
        \caption{Mean SHAP Values}
        \vspace{-1mm}
        \resizebox{\columnwidth}{!}{%
            \begin{tabular}{|lccc|}
                \toprule
                \textbf{RCP} & \textbf{Throughput} & \textbf{Spectral Efficiency} & \textbf{BLER}\\
                \midrule
                Tx Power            &0.5474     &0.0298     &0.226         \\
                Bandwidth           &0.0162     &0.0583     &0.004         \\
                PRBs                &0.0336     &0.0019     &0.0872         \\
                Num of Tx Antenna   &0.0499     &0.0026     &0.0175         \\
                \bottomrule
            \end{tabular}
        }%
        \label{tab:shapValues}
    \end{subtable}
    \label{tab:featureImpScores}
\end{table}

\begin{figure}[tb!]
\centering
    \begin{subfigure}[t!]{0.85\columnwidth}
         \centering
         \includegraphics[width=\textwidth]{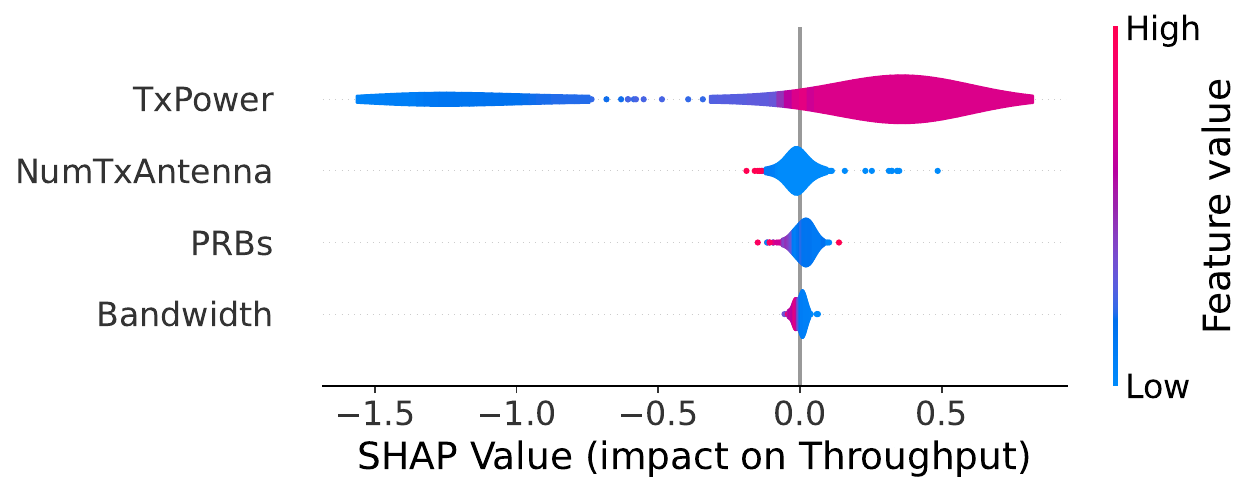}
         \caption{Throughput}
         \label{fig:shap1}
     \end{subfigure}
     \begin{subfigure}[t!]{0.85\columnwidth}
         \centering
         \includegraphics[width=\textwidth]{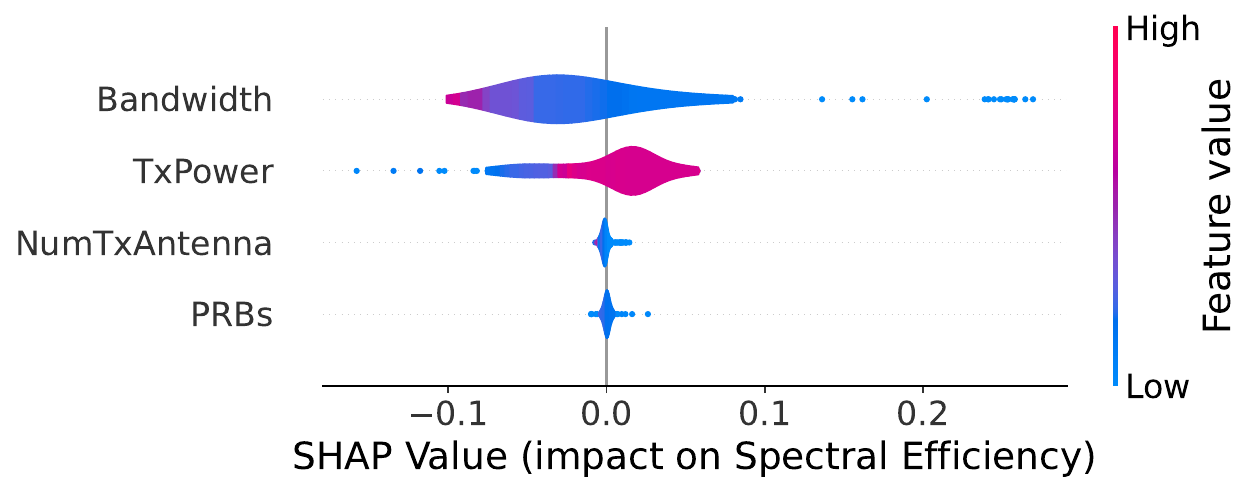}
         \caption{Spectral Efficiency}
         \label{fig:shap2}
     \end{subfigure}
     \begin{subfigure}[t!]{0.85\columnwidth}
         \centering
         \includegraphics[width=\textwidth]{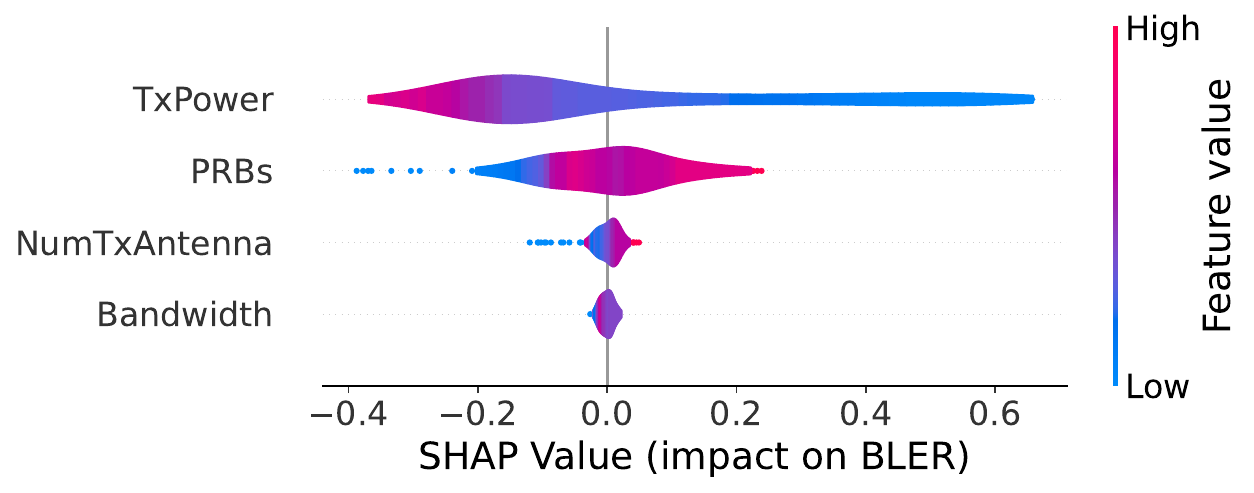}
         \caption{BLER}
         \label{fig:shap3}
     \end{subfigure}
\caption{SHAP value plots indicating the feature importance of each RCP on KPIs (a) Throughput, (b) Spectral Efficiency, and (c) Block Error Rate (BLER)}
\label{fig:shap}
\end{figure}
From the above experiments, we obtain a tabular dataset with 7 columns indicating values of different RCPs and KPIs and 1000 rows representing each simulation episode. 
We perform regression of KPI values over all the RCPs using a range of ML models:  Decision Trees, Random Forest, XGBoost, Multi-Layer Perceptron (MLP), and Support Vector Regression (SVR). The predictive performance of these models is evaluated using the coefficient of determination (R\textsuperscript{2}), Mean Squared Error (MSE), and Mean Absolute Error (MAE), as summarized in Table~\ref{tab:regression_metrics}. Higher R\textsuperscript{2} values (closer to 1) indicate a better model fit, while lower MSE and MAE values correspond to greater model accuracy. 
The results demonstrate that tree-based ensemble methods, particularly Random Forest and XGBoost, outperform MLP and SVR across all KPIs, with XGBoost exhibiting the best overall performance.

To answer the \textbf{RQ1}, we analyze feature importance of RCPs on each KPI using permutation importance scores and SHAP values as mentioned in Table~\ref{tab:featureImpScores}. We choose XGBoost to obtain these values because of its highest performance in regression. 
As seen in Figure~\ref{fig:shap}, RCPs are ranked according to their overall importance in the SHAP value plots. SHAP feature importance plots also indicate the direction of movement of KPIs when the RCPs are increased or decreased. From the feature importance scores and the SHAP plots, we can observe that Tx power is the most important RCP influencing throughput.
Bandwidth and Tx power both affect spectral efficiency. Tx power and number of PRBs both influence BLER, however, Tx power is the more dominant RCP. Although it seems counterintuitive that either the bandwidth or PRBs is not reflecting importance in affecting throughput, we argue that the effect of Tx power in our simulated data is significantly more than the other RCPs, which masks the importance of bandwidth or PRBs one would expect on throughput. Future investigations on data from real xApp deployments in O-RAN testbeds may eliminate such irregularities. Regardless, we use this evaluation to demonstrate how feature importance tools like SHAP can be used to identify potential conflicts.

\subsection{Conflict Effect Evaluation via Causal Inference}
\begin{figure}[tb!]
    \centering
    \includegraphics[width=0.65\columnwidth,trim={0cm 0.5cm 10cm 0.7cm},clip]{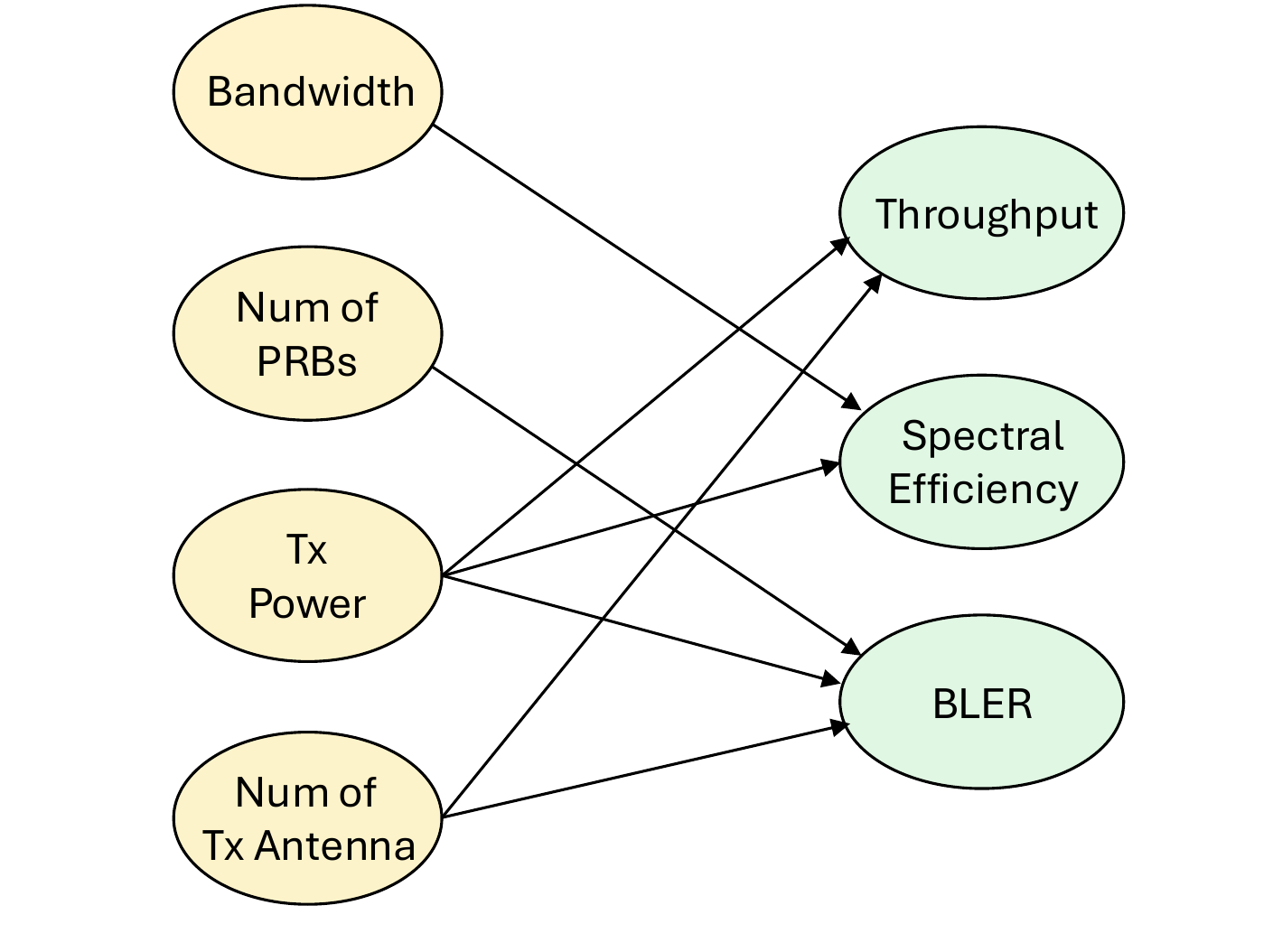}
    \caption{Causal Directed Acyclic Graph (DAG)}
    \label{fig:causaldag}    
\end{figure}

We construct a causal \gls{dag}, as shown in Figure~\ref{fig:causaldag}, based on the SHAP analysis performed in the previous step. This DAG captures the dependency structure between RCPs and KPIs. To address \textbf{RQ2}, we apply a formal causal analysis framework using the DoWhy~\cite{dowhy} and EconML~\cite{econml} libraries.
We analyze treatment-outcome pairs identified in the DAG and estimate the Average Treatment Effects (ATEs), which quantify the causal influence of RCPs on KPIs. These estimates, summarized in Table~\ref{tab:ateValues}, correspond to the causal edges shown in Figure~\ref{fig:causaldag}.
For instance, we observe that a 1dBm increase in Tx power (an RCP associated with a power allocation xApp) yields an average increase of 0.00251 in spectral efficiency. Conversely, a 1 MHz increase in bandwidth (RCP of a spectrum optimization xApp) results in an average decrease of 0.00320 in spectral efficiency, as shown in Table~\ref{tab:ateValues}. These ATE values enable the MNOs to design a conflict mitigation framework that can notify the xApps of the overall impact of their actions and suggest parameter adjustments to remain within acceptable operational tolerances. 

\begin{table}[b!]
    \centering
    \caption{ATEs and p-values from refutation tests for Treatment-Outcome pairs identified from the causal DAG}
    \resizebox{\columnwidth}{!}{%
    \begin{tabular}{|llcc|}
    \toprule
         \textbf{Treatment (RCP)} & \textbf{Outcome (KPI)} & \textbf{ATE} & \textbf{p-value}\\
         \midrule
         Tx Power & Throughput & 0.04449 & 0.88\\
         Tx Power & Spectral Efficiency & 0.00251 & 0.96\\
         Tx Power & BLER & -0.01816 & 0.86\\
         Bandwidth & Spectral Efficiency & -0.00320 & 0.94\\
         Num of PRBs & BLER & 0.00180 & 0.98 \\
         Num of Tx Antenna & Throughput & -0.00834 & 0.92\\
         Num of Tx Antenna & BLER & 0.00156 & 0.90\\
         \bottomrule
    \end{tabular}
    }%
    
    \label{tab:ateValues}
\end{table}

To validate the robustness of the \gls{ate} estimates, we conduct three standard refutation tests: (i) Placebo Treatment Test, where the treatment variable is replaced with an unrelated random variable to rule out spurious correlations; (ii) Random Common Cause Test, which introduces synthetic confounders to ensure the estimated effect is not sensitive to noise; and (iii) Data Subset Test, where the estimation is repeated on random subsets of the original data to confirm the stability and generalization of our estimates. Each of these tests yielded high p-values ($>$ 0.05), confirming the statistical robustness of the estimated causal effects. Here, the p-value represents the probability of observing the estimated effect under the null hypothesis of no causal relationship. Table~\ref{tab:ateValues} includes p-values obtained from the Random Common Cause Test.

\begin{figure}[tb!]
    \centering
    \begin{subfigure}[t!]{\columnwidth}
        \includegraphics[width=\textwidth]{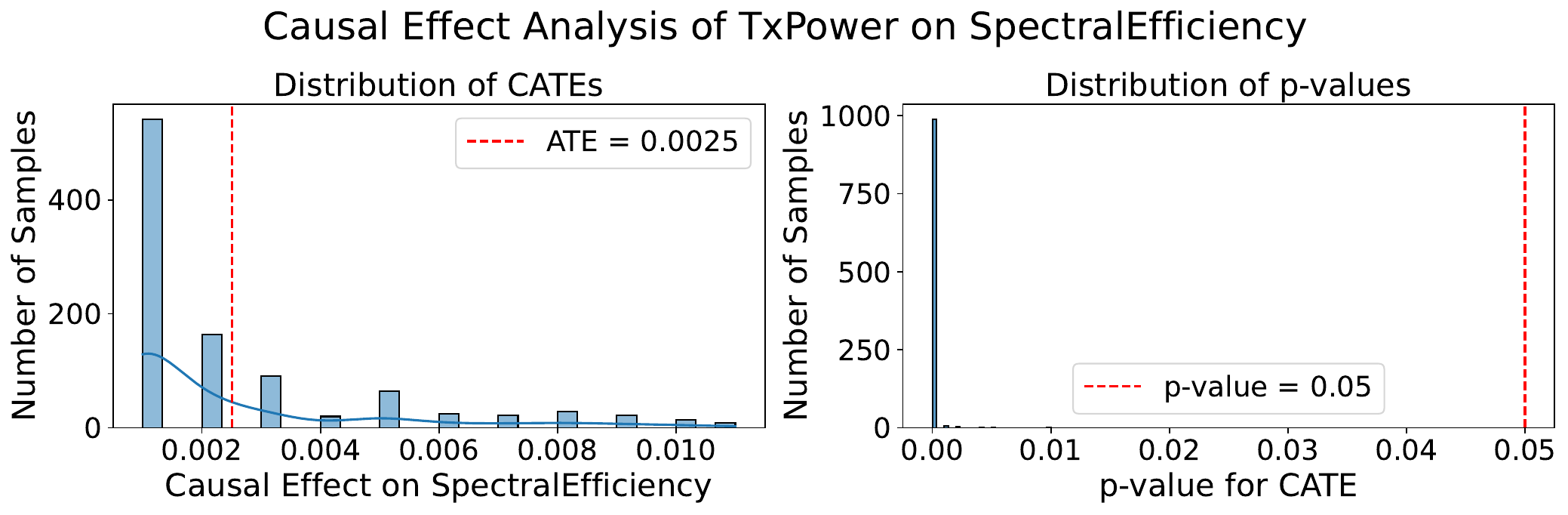}
        \label{fig:cate_txpower}
    \end{subfigure}
    
    \begin{subfigure}[t!]{\columnwidth}
        \includegraphics[width=\textwidth]{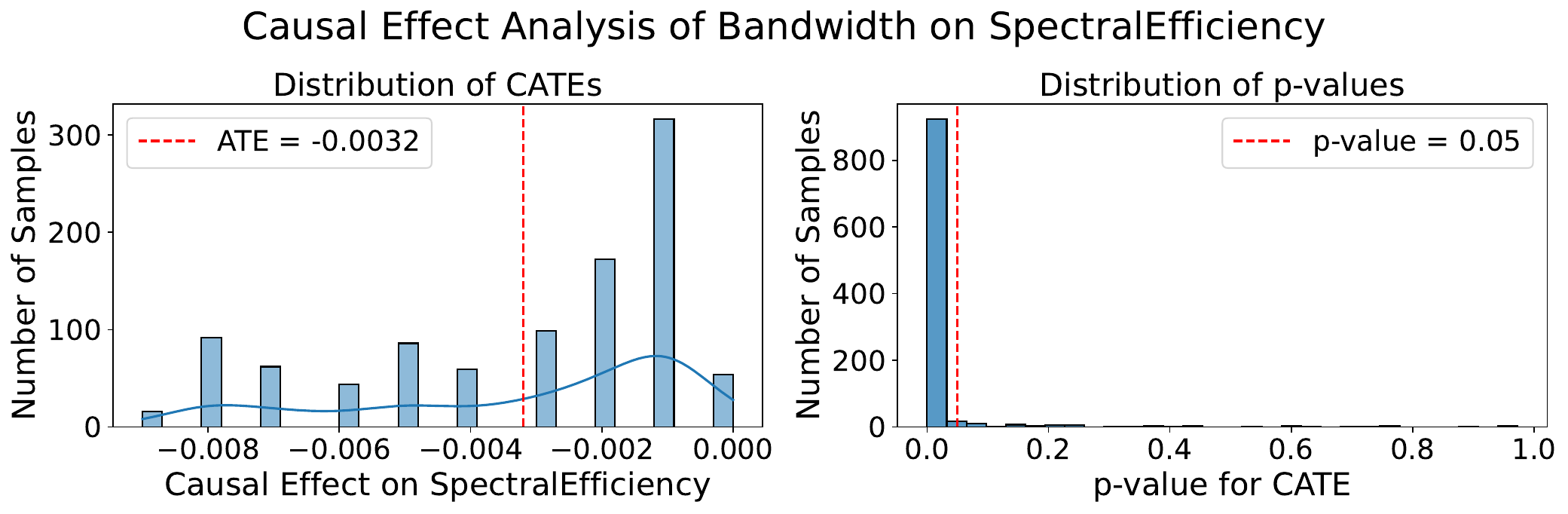}
        \label{fig:cate_bw}
    \end{subfigure}
    
    
    
    \vspace{-2mm}
    \caption{CATEs of Tx Power (top) and Bandwidth (bottom) on Spectral Efficiency}
    \label{fig:cates}
\end{figure}

We compute \acrfull{cate} values for the treatment–outcome pairs listed in Table~\ref{tab:ateValues} using the CausalForestDML model from the EconML package, with XGBoost selected as the base regressor due to its superior regression performance (as shown in Table~\ref{tab:regression_metrics}). Figure~\ref{fig:cates} shows the distribution of CATEs and corresponding p-values for two RCP-KPI pairs from the causal DAG (others omitted for the sake of brevity). The p-value distributions are tightly centered well below the 0.05 significance threshold, indicating strong statistical significance across a wide range of network states. Furthermore, we validate the robustness of our findings by conducting a sensitivity analysis by re-estimating the causal effects using the LinearDML model with polynomial features to assess consistency across different estimation techniques.

As observed in Figure~\ref{fig:cates}, while an increase in Tx power increases the overall spectral efficiency (positive ATE), the magnitude of the effect varies across the data, with CATE $\approx$ 0.001 for most samples. Similarly, the effect of increased bandwidth is a decrease in spectral efficiency with several CATE values $\approx$ –0.001. This fine-grained characterization of effect heterogeneity enables MNOs to move beyond a one-size-fits-all control policy, allowing for adaptive conflict resolution strategies tailored to specific network conditions.


\section{Limitations and Future Work}\label{sec:discussion}
The primary challenge in xApp conflict management is collecting a comprehensive dataset that captures the complexities of xApp actions (RCPs) and the corresponding network responses (KPIs). Obtaining such representative data is inherently difficult, and the lack of publicly available datasets restricts the evaluation of our proposed approach to data generated from a simplified network simulation. We plan to enhance our work in the future by conducting experiments on an O-RAN simulator, such as ns-O-RAN~\cite{nsORAN}, with multiple xApp deployments to generate more realistic high-dimensional datasets and to evaluate the scalability of our approach.

\section{Conclusions}\label{sec:conclusion}

This work addresses xApp conflict management by causally analyzing the relationships between xApp control parameters (RCPs) and network performance metrics (KPIs). We use ML model explainability methods like SHAP, to interpret a regression model trained on network data. This helps us to identify the RCPs that influence the same KPI, which can lead to potential conflicts. Subsequently, we leverage the SHAP analysis to construct a causal Directed Acyclic Graph (DAG) and estimate both the average (ATE) and conditional (CATE) treatment effects of these conflict-causing RCPs on KPIs. This combination of explainable ML and causal inference allows for a fine-grained analysis of xApp conflict impacts, guiding the design of more effective conflict resolution policies.

\section*{Acknowledgment}
We thank the anonymous reviewers for their valuable feedback and Joao Santos of the Commonwealth Cyber Initiative (CCI) at Virginia Tech for his insightful comments on the initial draft of this manuscript. This work was supported by the U.S. National Science Foundation (NSF) under Grant Number CNS-2317829 and the U.S. Defense Advanced Research Projects Agency (DARPA) under agreement number HR001120C0154.  
The views, opinions, and findings in this article are those of the authors and should not be interpreted as representing the official views or policies, either expressed or implied, of the NSF, DARPA or the Department of Defense.

\bibliographystyle{IEEEtran}

\end{document}